# A Novel Implementation for Automated Health Monitoring System

Praveen B Sarangamath[1], Dr. Kiran A Gupta[2]

[1]M.Tech. Dept of E&CE, Dayananda Sagar College, Bangalore, Karnataka, India
[2] Prof. and Head Dept of Medical Electronics, Dayananda Sagar College, Bangalore, Karnataka, India

*Abstract*— As population is increasing and ageing, there is also increase in the chronic and heart diseases. The current hospital centric healthcare is becoming inefficient to treat the conditions which demand immediate treatment such as heart strokes. So, the focus is now tilting from hospital centric treatment to patient centric treatment [7]. This paper proposes an idea of health monitoring system which monitors the physiological parameters of the patient such as temperature and heart beat rate using sensors which are connected to XBEE module that transmits the information to the local server and also log the collected information into the server database placed in the range of 20 meters.. A GSM module is incorporated into the project which alerts the doctor if any vital parameter of the patient deviates from the normal value anywhere through SMS. It helps the doctor to monitor the patients health parameters such as temperature, heart beat rate, blood pressure, moisture detection etc. The main concept of the proposed work is derived from Wireless Body Area network (WBAN). The proposed work employs Raspberry Pi kit as a personal server which logs the health data and it can be accessed by any PDA within the LAN range. In this paper, two vital parameters namely Temperature sensor and Heart beat sensor have been considered.

## I. INTRODUCTION

The concept of automated health monitoring system is taken from Wireless Body Area Network (WBAN). The proposed system model assists the doctor in analysing the health record of the patient and helps in proper diagnosis. A lot of researches are taking place in the wearable sensor field in recent years. Wearable sensors are in contact with the human body and monitor his or her physiological parameters. One can buy variety of sensors in the market today such as pulse monitors, oximetry monitors, moisture monitors etc. The cost of the sensors varies according to their size, flexibility and accuracy. Some literature surveys related to the work are discussed here.

Chin Teng Lin and others [2] presented the idea of dry EEG sensor based mobile wireless system which aims at recording the EEG signals and evaluation of driver vigilance. Grantham K.H. Pang [4] proposed the idea of Responsive health monitoring system which provides automated data analysis and response to the collected data. However this idea requires the human intervention to upload the data into the web portal. Purnima and Puneet singh [1] proposed an idea of Zigbee based health monitoring system in which the patients vital parameters are continuously measured and the data is sent to the centralised ARM controller for data analysis. The data so collected is updated to the doctor PC via Zigbee receiver module. The range of coverage is less since the presented idea covers only few meters. Another drawback of the system is that it is using less calibrated heart beat sensor. In the proposed idea the author is aiming at improving the range of coverage to LAN from personal Area Network (PAN) by building a personal server and the work is also employing highly accurate and low cost heart beat sensor from Sparkfun Corporation and Li2 Technologies.

## II. SYSTEM MODEL

The proposed system model is described using block diagram and flow chart. The block diagram as shown in the Figure 1 has three different phases namely data acquisition phase, data transmission phase and data displaying phase. In data acquisition phase the vital parameters of the patient are gathered using the physiological sensors. In this proposed idea temperature sensor and heart beat sensors are employed. In data transmission phase the collected data is sent to the local server using Zigbee technology. In the final phase i.e. data display phase a web page is built and the data acquired is displayed on the web page using raspberry pi which is acting as local server.

A. *Block Diagram*

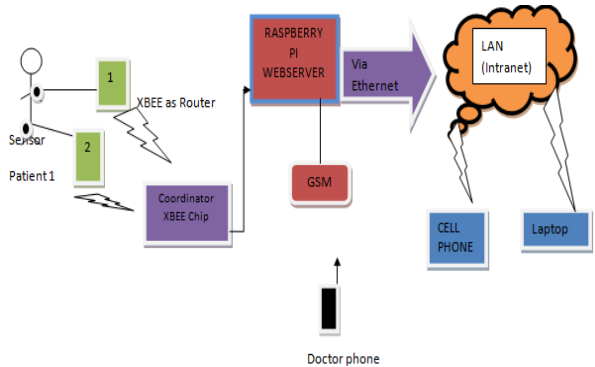

Figure 1: Block Diagram of System

B. *Flow chart*

The Figure 2 describes the flow chart of the proposed system. Initially temperature sensor and heart beat sensors capture the data from the patient and are compared against the normal values. If the values exceed the normal range the GSM module is activated and SMS is sent to the doctor. Otherwise the data captured is simply stored in the local server and the values are displayed on the web page upon request.

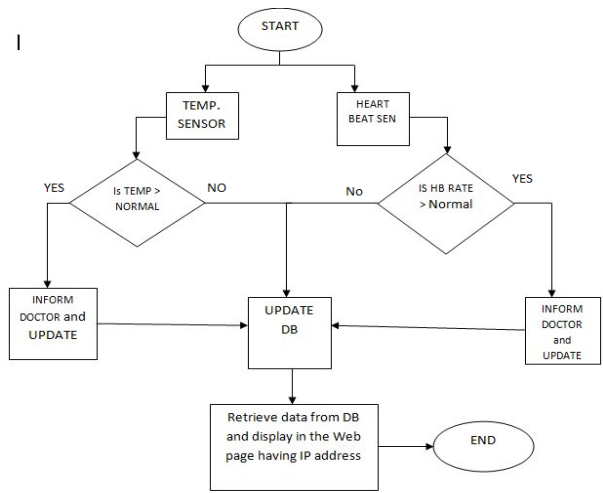

Figure 2: Flow chart of the system

### III. DESIGN METHODOLOGY

The system design is mainly organised into two parts namely Hardware interfaces and hardware, software details. The hardware and software used in the system are discussed in detail in this section.

A. *Hardware Interfaces:*

Figure 3 describes the connection between Raspberry Pi and XBEE coordinator chip. Raspberry Pi model B+ is 40 pin chip as said earlier. The first pin of the Raspberry Pi (3.3V pin) connected to 3.3V pin of XBEE coordinator which is the first pin on XBEE. The second pin i.e. transmitter pin of XBEE is connected to the tenth pin of the Raspberry Pi. The tenth pin of the XBEE (i.e. the ground pin) is connected to sixth pin of Raspberry Pi which is ground pin.

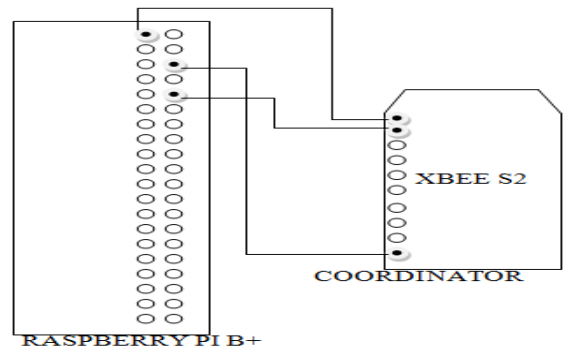

Figure 3: Raspberry Pi and XBEE connection

Figure 4 describes the connections between Raspberry Pi and SIM 900 GSM modem.

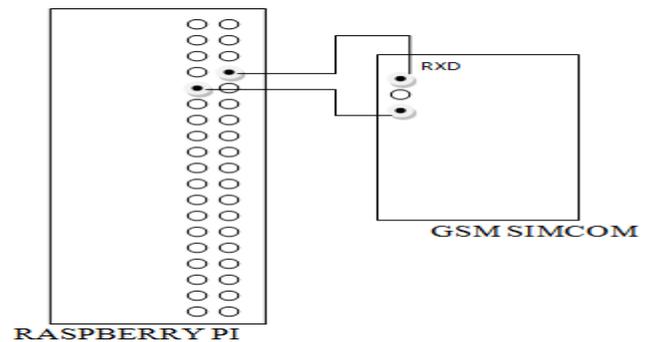

Figure 4: Raspberry Pi and GSM modem interface

SIM900 modem has three pins namely RXD pin, TXD pin and ground pin. Along with this, the module has a power supply port to power up the GSM module. The eighth pin of Raspberry Pi that is the Transmitter pin is connected to the receiver pin (RXD) of SIMCOM900. The ninth pin (i.e. Ground) of Raspberry Pi is connected to the ground pin (GND) of GSM modem. The transmitter pin is left open, since this pin is not used.

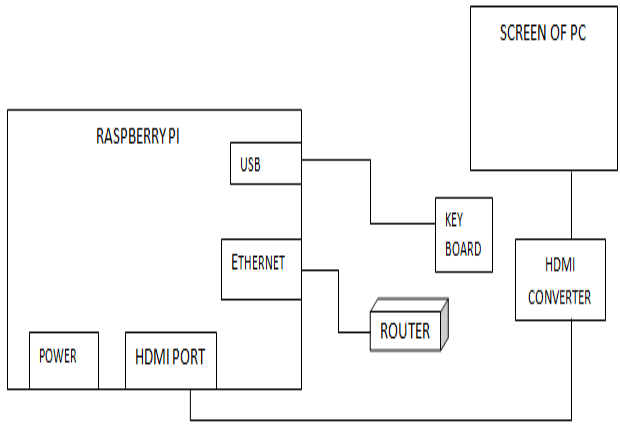

Figure 5: Raspberry Pi to monitor connections

The display screen (Monitor) of PC is connected to the HDMI converter if the PC does not support HD monitor. The output of HDMI converter is connected to the HDMI port of Raspberry Pi. If the PC has HD monitor then no need of HDMI converter. An Ethernet cable is used to interface router to the Ethernet port of Raspberry Pi. The power port is used to provide power supply to the Raspberry Pi.

### B. Hardware Details

1) *XBEE radio module*: XBEE is the radio module launched by Digi. This works on 802.15.4 standard. There are two flavours of XBEE series namely S1 and S2. This idea employs XBEE S2 over S1 because, the range of coverage for S2 is more than S1 and the output power is also more in S2 than S1. XBEE S2 is a twenty pin chip which operates at 3.3V power supply and it functions in two modes namely API (Application Programming Interface) mode and AT (Attention) Mode [8].

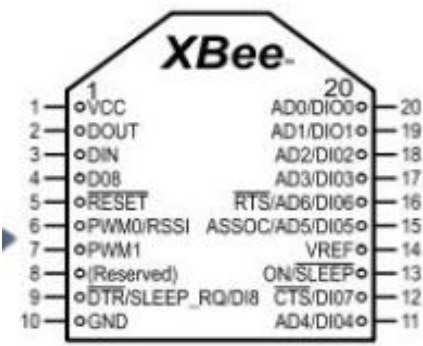

Figure 6: Pin Configuration of XBEE S2

2) *Temperature Sensor*: LM35DZ sensor is employed in the implementation to collect the temperature information from the patient. LM35 is more accurate sensor with an accuracy of +/- 0.4C. It has a low self heating capability and draws only 60 micro amps of current.

3) *Heart beat count sensor*: AD 8232 and reflectance based sensor from Li2 Technologies are used to capture heart beat graph and heart beat counts respectively. The heart beat count sensor from Li2 Technologies is shown in Figure 7.

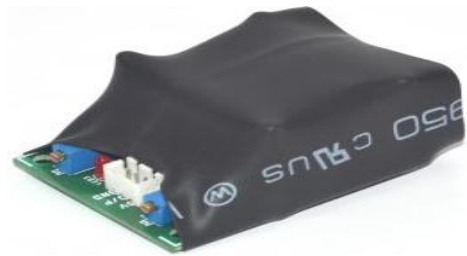

Figure 7: Heart beat count sensor

The finger tip is placed on the IC. After some time the IC synchronises with the heart beats of human and starts giving out the heart beats per minute. It uses one light emitting diode and one detector which capture the light reflected from the finger. It calculates the intensity of the received light and based on this the heart beats are measured.

4) *Raspberry Pi*: Raspberry Pi is a credit card sized CPU developed by Raspberry Foundations in UK. Raspberry Pi kit includes an ARM1176JZF-S700 MHz processor, Video Core IV GPU, and was originally comes with 256 megabytes of RAM. In the improved versions RAM size is extended. It supports different programming platforms like java, python, Perl, ruby etc. In this work author is using Raspberry Pi to build a personal server which stores the data in a database. The kit consists of four USB ports, one Ethernet port and a power port. There are three versions of raspberry at present namely model A, B and B+. The proposed work is employing Raspberry Pi B+ model.

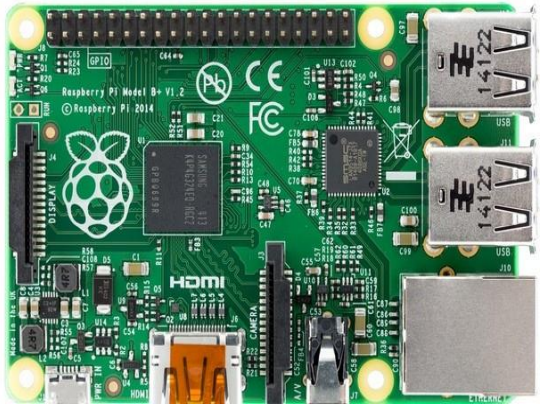

Figure 8: Raspberry Pi model B+

5) *GSM modem*: This modem is used to send SMS to the concerned doctor if any of the parameters exceeds normal value. This idea employs SIM900 as GSM modem.

C. *Software Details*

1) *X-CTU Tool:* This is XBEE chip configuration tool given by Digi internationals. Through this software utility tool XBEE chips *can be* configure as Coordinator, router or end devices. One chip can be configured as transmitter and other as receiver to set up a simple Zigbee network.

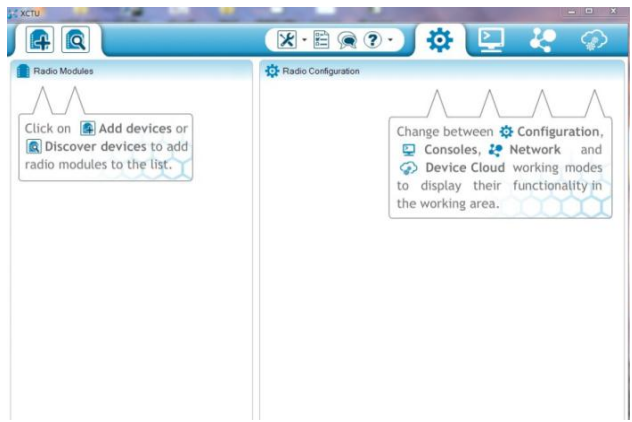

Figure 9: X-CTU NG view

2) *Python programming environment*: Raspberry Pi is configured as a personal web server using python programming. The reason for using Python is that it is an interpreted language and it has very large library in terms of Terabytes. It is easy to maintain the code and it provides interface to all commercial data bases. So, python platform is chosen over other programming platforms.

IV. RESULTS

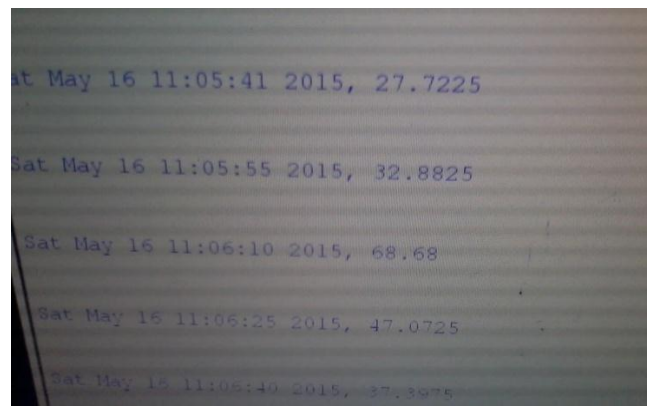

Figure 10: Temperature results

The Figure 10 shows the temperature values captured from the body. It stores the temperature values in degree Celsius along with the time and date.

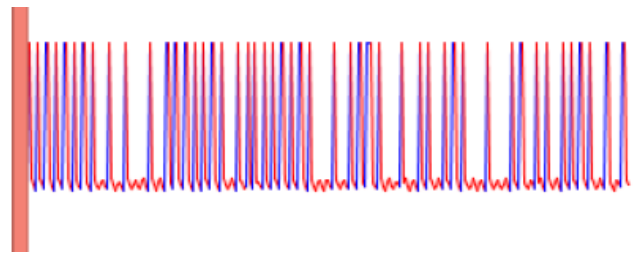

Figure 11: Heart beat graph obtained from AD8232

The Figure 11 depicts the electrical signals generated from heart. These signals are captured by AD8232 sensor and are displayed on the screen.

Figure 12 shows the front end web page view. This web page displays the variation of temperature in the form of graph and also the current heart beats in terms of per minute.

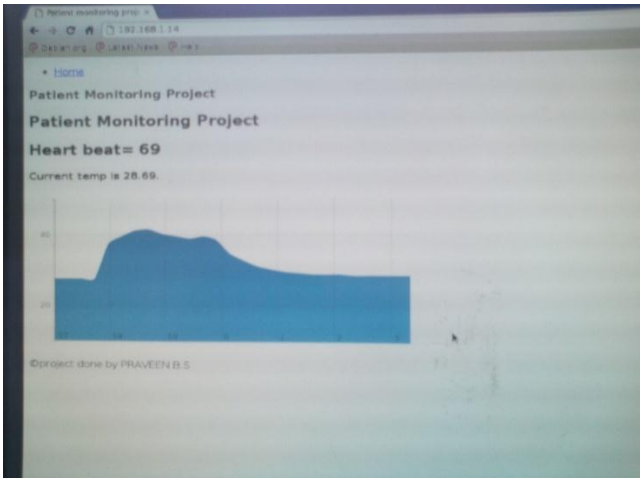

**Figure 12: Front end web page**

## V. WORKING METHODOLOGY

Initially the sensors are placed on the patient body at the strategic places. Depending upon the sensors used the data corresponding to that physiological parameter is captured by the sensors which are given to microcontroller for data analysis. The processed data is sent to the XBEE chip which is acting as a wireless module to transmit data. The data sent from the XBEE router is received at the XBEE coordinator and the data is subsequently stored in the data base by Raspberry Pi. Raspberry is pre configured to store the data. As a next stage Raspberry is configured as personal web server which can be accessed by any devices that gets connected to that LAN. A web interface is developed which provides user interface to display the data.

## VI. APPLICATIONS

This proposed idea is based on ZIGBEE technology which is more cost effective compared to Bluetooth. The range of communication cab be improved by developing ZIGBEE networks. This proposed idea aims at using a low cost but accurate sensors which monitors the vital parameters of the patient more accurately thus assisting the doctor in the better diagnosis.

## VII. CONCLUSION AND FUTURE SCOPE

The health monitoring system proposed in this paper is developed with the aim of providing the doctors the much needed patient health history in real time.

This will assist the physicians in proper diagnosis and treatment. This topic is one of the hot cake for researchers. The Apple Corporation has launched a watch which can sense the heart beat and display it in the UI screen. The heart beat sensor employed in this paper works almost on the same principle employed by Apple watches in measuring the heart beats.

There is always a scope of development in this field since this is highly evolving field. This paper proposes the use of temperature and heart beat sensors, but this can be extended to measure other vital parameters such as EEG, pulse oximetry, Blood pressure etc. Hosting services from go daddy or other organizations can be used to store the data in global server, thus removing the barrier of range constraint.

*Acknowledgement*

I would like to thank my guide Dr. Kiran A Gupta, Prof and Head of Medical Electronics for her valuable inputs at all stages of the work. I would like to extend my regards to Dr. G.V. Attimarad, Prof and Head of Electronics and Communication for inspiring us to undertake research works in the field of Electronics.

REFERENCES

[1] Purnima, Puneet singh, 2014 ,Zigbee and GSM based Patient and health Monitoring system. In International Conference on Electronics and Communication systems ICECS.

[2] Chin-Teng Lin, Chun-Hsiang Chuang, Chih-Sheng Huang, Shu-Fang Tsai, 2014,Wireless and Wearable EEG System for Evaluating Driver Vigilance. IEEE TRANSACTIONS ON BIOMEDICALCIRCUITS AND SYSTEMS, VOL. 8, NO. 2.

[3] Sarkar Suvojit and Das Rajdip, 2013, An autonomous wireless hospital management system. Research Journal of Engineering Sciences, Vol2 (9), 7-9.

[4] Grantham K.H. Pang, 2012, Health monitoring of Elderly in Independent in Assisted Living. In International Conference on Biomedical Engineering ICoBE, 27-28.

[5] Nisha Singh and Sr.Asst.Prof. Ravi Mishra, Microcontroller based wireless temperature and heart beat read out", IOSRJEN, Vol.3 PP-01-06,Jan .2013.

[6] Carles Gomez and Josep paradells, "Wireless Home Automation Networks: A survey of Architectures and Technologies", IEEE Communications Magazine, Jun 2010.. Mullende

[7] Pedro Brandao, "Abstracting information on body area networks", University of Cambridge , ISSN: 1476-2986.

[8] www.digi.org, a web resource.